\documentclass[twocolumn]{article}
\usepackage{graphicx}
\usepackage{latexsym}
\usepackage{amssymb,amsmath}
\usepackage{epsfig}

\title{Cosmological consequences of the redistribution of energy between matter components in the very early universe}
\author{V.E. Kuzmichev, V.V. Kuzmichev}
\date{Bogolyubov Institute for Theoretical Physics,\\
National Academy of Sciences of Ukraine,\\
Metrolohichna Str. 14-B, 03680 Kiev, Ukraine}

\begin{document}



\maketitle

\begin{abstract}
The evolution of matter in the expanding FRW universe during the time interval between the end of 
inflation and the beginning of the radiation-dominated era is studied.
A constraint between the global geometry and total amount of matter in the universe as a whole,
which is valid during the phase of an intensive transfer of energy to the matter degrees of freedom, is introduced.
The matter is considered as a perfect fluid with two components between which there is energy exchange.
The analytical solutions of the Einstein equations are found.
The limiting cases of the the Hubble expansion rate and the total energy density, which
correspond to matter production, pressure-free and radiation-dominated phases are investigated.
The transition to the inflationary phase and a unidirectional evolution of matter in the 
universe at all phases are discussed.
\end{abstract}

\section{Introduction}
The standard cosmological model is based on the hot Big Bang model and the assumption about the inflationary expansion 
of the very early universe, when the scale factor grows quasi-exponentially, while the Hubble expansion rate remains almost constant
(e.g., Refs. \cite{Lin,Dol,Ly,Rio}). 
During inflation the universe is in the vacuum-like state which is usually associated with a scalar field called the inflaton. 
After inflation the energy density of the primordial matter (except the inflaton) which filled the universe before this stage 
becomes negligibly small. In order to explain the presence of conventional matter in the universe after inflation, the decay of 
the vacuum-like state into `normal' particles is postulated. 

The universe becomes hot as a result of interaction between particles and
transits into the radiation-dominated phase. In the process of energy transfer from the inflaton to radiation (called reheating) 
the equation of state of matter changes. A change from a vacuum-like equation of state to the equation of relativistic matter might be gradual
and an intermediate stage between these two known phases may be modeled.

In the present article, the evolution of the equation of state of the matter in the universe during the time interval between the end of 
inflation and the beginning of the radiation-dominated era is considered. Without rendering concrete mechanisms of decay of vacuum-like state 
into the conventional matter, we assume that the global geometry and total amount of matter in the universe as a whole satisfy a constraint, 
which is valid during some time interval, before radiation domination.
This constraint is equivalent to the law of the conservation of total energy of the universe which remains equal to zero due to the 
gravitational mass effect, whereas the energy attributed to the particles of conventional matter increases with expansion of the universe 
\cite{Dol,Zel}. 
In this case, at all stages of evolution the universe is described by the Einstein equations with addition of appropriate equations of state.

The paper is organized as follows. In Section 2 the basic equations which describe the homogeneous, isotropic and spatially flat universe 
are given. The equations of state of matter for the different phases of reheating are justified. In Section 3 a two-component perfect
fluid model is introduced. The analytical solution of the non-linear equation for the Hubble expansion rate is obtained. The expressions
for the deceleration parameter and the total energy density as functions of time are deduced. The limiting cases of the solutions which
correspond to pressure-free and relativistic matter are considered. The Whitrow-Randall's relation \cite{Whi} is rederived. In Section 4
the transition to the inflationary phase is discussed. The mechanical analogy which explains a unidirectional evolution of matter in the 
universe at the phases under consideration is given.

\section{Equation of state parameter}
Let us consider the homogeneous, isotropic and spatially flat universe in the early epoch, when its dynamics can be described by the equations of 
the Friedmann-Robertson-Walker (FRW) cosmology,
\begin{equation}\label{1}
H^{2} \equiv \left(\frac{\dot{R}}{R} \right)^{2} = \frac{8 \pi G}{3} \rho,
\end{equation}
\begin{equation}\label{2}
\dot{\rho} + 3 H (\rho + p) = 0,
\end{equation}
\begin{equation}\label{3}
p = w(t)\, \rho,
\end{equation}
where $R(t)$ is the cosmic scale factor, $\rho(t)$ is the energy density of the matter which has a form of the 
homogeneous perfect fluid, $p(t)$ is its pressure, $w(t)$ is the equation of state parameter,
$G$ is the Newtonian gravitational constant, an overdot denotes $d/dt$, $t$ is the proper time 
(units $c = 1$ are used). 

Let there exist an interval of time after inflation $\Delta t_{ci} = t_{c} - t_{i}$, where $t_{i}$ denotes the time at which inflation ends and 
$t_{c}$ stands for the time at which an intensive transfer of energy to the matter degrees of freedom ends.
We assume that during this interval the matter is produced, so that the following condition is fulfilled, at least in good approximation,
\begin{equation}\label{4}
M - G \frac{M^{2}}{R} = 0,
\end{equation}
where $M = \frac{4}{3} \pi R^{3} \rho$ is total mass-energy of matter (the sum of masses of particles of conventional matter)
in the equivalent flat-space volume taken without account of gravitational interaction between particles. 
The equation (\ref{4}) can be interpreted as the law of the conservation of zero total mass-energy of the universe 
during its expansion with matter production \cite{Zel}. 
According to Eq. (\ref{4}), during the time interval $\Delta t_{ci}$ the following relation $R = G M$ is valid. It means
that the energy density 
\begin{equation}\label{5}
\rho = \frac{3}{G} \frac{1}{4 \pi R^{2}}
\end{equation}
decreases linearly with increasing surface area $4 \pi R^{2}$.

From Eqs. (\ref{2}) and (\ref{5}), one can find the equation of state
\begin{equation}\label{6}
p = - \frac{1}{3} \rho, \quad w(t) = -\frac{1}{3}.
\end{equation}

After the end of this phase, the mass $M$ remains constant on the time interval $\Delta t_{rc} = t_{r} - t_{c}$, 
where $t_{r}$ denotes the beginning of the subsequent relativistic matter dominant era. 
The equation of state on the time interval $\Delta t_{rc}$ takes the form
\begin{equation}\label{7}
p = 0, \quad w(t) = 0.
\end{equation}

In the relativistic matter dominant era, the mass attributed to relativistic matter reduces as the universe expands, 
$M \sim R^{-1}$, due to the cosmic redshift. For times $t > t_{r}$, the equations of state has a form
\begin{equation}\label{8}
p = \frac{1}{3} \rho, \quad w(t) = \frac{1}{3}.
\end{equation}

We will study the model of evolution of matter in the early universe, 
where the equation of state parameter $w(t)$ changes with time from $-\frac{1}{3}$,
passing through the point $w = 0$, to $\frac{1}{3}$ taking all intermediate values. 
Substituting a continuous function $w(t)$ of time $t$,
\begin{equation}\label{9}
w(t) = \frac{1}{3} \tanh \left(\frac{t - t_{0}}{\tau}\right),
\end{equation}
for the equation of state parameter on the time interval 
$[t_{i}, t_{r}[$, and choosing properly a point $t_{0}$ on this interval, 
one can reproduce the required values of $w$ (\ref{6})-(\ref{8}).
Since $t_{r} \gg t_{i}$ (in standard cosmological model, the value $t_{i} \sim 10^{-35}$ s is acceptable, whereas the time $t_{r}$ is often evaluated as $t_{r} \sim 10^{-30}$ s corresponding to temperatures not exceeding
$10^{12}$ GeV \cite{Lin}), the good estimation for $t_{0}$ may be $t_{0} \lesssim t_{r}$.
The value $1/\tau$  determines the mean rate of change of the equation of state parameter $w(t)$. 
Such a variation of the equation of state parameter can be achieved in a system, where the matter consists of a few components
between which occurs the energy transfer for some typical time $\tau$.

\section{Two-component fluid}
Let us consider a two-component perfect fluid with the energy density and pressure
\begin{equation}\label{10}
\rho = \rho_{q} + \rho_{d},\qquad p = p_{q} + p_{d}.
\end{equation}
These components satisfy the equations
\begin{equation}\label{11}
\dot{\rho}_{q} + 3 H (\rho_{q} + p_{q}) = Q,\quad \dot{\rho}_{d} + 3 H (\rho_{d} + p_{d}) = -Q,
\end{equation}
which represent the energy conservation law (\ref{2}) rewritten for components,
$Q$ is the interaction term.

The components of the perfect fluid are imitated by scalar fields $\phi_{q} (t)$ and $\phi_{d} (t)$
with potentials $V_{q}(\phi_{q})$ and $V_{d}(\phi_{d})$,
\begin{equation}\label{12}
\rho_{\alpha} = \frac{1}{2} \dot{\phi_{\alpha}}^{2} + V_{\alpha},\ p_{\alpha} = \frac{1}{2} \dot{\phi_{\alpha}}^{2} - V_{\alpha},
\ \alpha = \{q,d\}.
\end{equation}
The models of such a type which include a coupling between the matter components
were considered in the literature, in particular, within the context of inflation and reheating
and the coincidence problem of dark energy and matter in the accelerating universe 
(see, e.g., Refs. \cite{Bil,Ame,Zim} and references therein). 
The form of the interaction term $Q$ may be derived from different physical arguments or obtained as a solution 
of some dynamical equation, which describes the required properties of the matter fields $\phi_{\alpha}$.

Let us assume that the field $\phi_{d}$ forms the pressure-free matter component (dust),
\begin{equation}\label{13}
\frac{1}{2} \dot{\phi_{d}}^{2} = V_{d},\quad \rho_{d} = 2V_{d},\quad p_{d} = 0.
\end{equation}
Concerning the field $\phi_{q}$, we suppose that it is described by the vacuum-type equation of state (as for the inflaton)
at times $t \ll t_{0}$,
\begin{equation}\label{14}
p_{q} \simeq - \rho_{q}.
\end{equation}
From Eq. (\ref{12}), it follows that at this stage the kinetic energy of the field $\phi_{q}$ can be neglected
and the total energy is determined by its potential term,
\begin{equation}\label{15}
\dot{\phi_{q}}^{2} \simeq 0, \quad \rho_{q} \simeq V_{q}.
\end{equation}

For times $t \simeq t_{0}$, the equation of state takes the form
\begin{equation}\label{16}
p_{q} \simeq 0.
\end{equation}
It means that
\begin{equation}\label{17}
\frac{1}{2} \dot{\phi_{q}}^{2} \simeq V_{q}, \quad \rho_{q} \simeq 2V_{q}.
\end{equation}
Then, for the times $t \gg t_{0}$, the field $\phi_{q}$ describes the matter component with the energy density which is
almost equal to its kinetic energy,
\begin{equation}\label{18}
\rho_{q} \simeq \frac{1}{2} \dot{\phi_{q}}^{2},\quad V_{q} \simeq 0.
\end{equation}
This phase corresponds to the reheating of the pressure-free matter and provides the passage to relativistic matter domination.
The field $\phi_{q}$ here has a form of the stiff Zel'dovich matter,
\begin{equation}\label{19}
p_{q} \simeq \rho_{q} \quad \mbox{at} \ t \gg t_{0}.
\end{equation}
The continuous transition from Eq. (\ref{14}) to (\ref{16}), and then from (\ref{16}) to (\ref{19}) can be achieved if
the following condition is imposed on the field $\phi_{q}$
\begin{equation}\label{20}
\frac{1}{2} \dot{\phi_{q}}^{2} \, \mbox{e}^{- 2 \left(t - t_{0}\right) / \tau} = V_{q},
\end{equation}
where $\tau < \frac{1}{2} t_{0}$.

Taking into account Eqs. (\ref{10}), (\ref{12}), (\ref{13}), (\ref{15}), (\ref{20}), from Eq. (\ref{3}) we find
\begin{equation}\label{21}
w(t) =\frac{\mbox{e}^{2 \left(t - t_{0}\right) / \tau} - 1}{\mbox{e}^{2 \left(t - t_{0}\right) / \tau} + 1 + 2 V_{d}/V_{q}}.
\end{equation}
This relation passes into Eq. (\ref{9}), if one introduces the following additional condition on $V_{d}$,
\begin{equation}\label{22}
V_{d} = \rho_{q} = V_{q} \left[\mbox{e}^{2 \left(t - t_{0}\right) / \tau} + 1\right].
\end{equation}
Then from Eqs. (\ref{10}) and (\ref{13}), we get
\begin{equation}\label{23}
\rho = 3 \rho_{q}, \quad p = p_{q}, \quad w = \frac{p_{q}}{3 \rho_{q}}.
\end{equation}
In this case, the interaction term $Q = 2 H p_{q}$ and the set of equations (\ref{11}) reduces to one equation
\begin{equation}\label{24}
\dot{\rho}_{q} + 3 H \left(\rho_{q} + \frac{1}{3} p_{q}\right) = 0.
\end{equation}
From Eqs. (\ref{1}), (\ref{9}), and (\ref{24}), it follows the non-linear equation for the Hubble expansion rate,
\begin{equation}\label{25}
\dot{H} + \frac{1}{2} \left\{3 + \tanh \left(\frac{t - t_{0}}{\tau}\right)\right\} H^{2} = 0.
\end{equation}
The general solution of this equation is
\begin{equation}\label{26}
H (t) = \frac{2}{D(t)},
\end{equation}
where we denote
\begin{equation}\label{27}
D(t) = C t_{0} + 3 t + \tau \ln \cosh \left( \frac{t - t_{0}}{\tau}\right),
\end{equation}
$C$ is a constant of integration.

The deceleration parameter, $q = -1 - \dot{H}/H^{2}$, is equal to
\begin{equation}\label{28}
q (t) = \frac{1}{2} \left\{1 + \tanh\left( \frac{t - t_{0}}{\tau}\right)\right\}.
\end{equation}
The deceleration parameter changes from the value $q = 0$ for the equation of state (\ref{6}), 
through the point $q = \frac{1}{2}$ for Eq. (\ref{7}),
to $q = 1$ for Eq. (\ref{8}). Thus, in the model under consideration, the expansion of the universe is decelerating 
on the whole time interval from the end of inflation to the beginning of the radiation-dominated era.

The total energy density is
\begin{equation}\label{29}
\rho (t) = \frac{3}{2 \pi G D(t)^{2}}.
\end{equation}
The limiting cases of
the solutions (\ref{26}) and (\ref{29}) reproduce the well-known expressions for the Hubble expansion rate and
the energy density. Setting $C = 0$, near the point $t = t_{0}$ we find for pressure-free matter \cite{LL}
\begin{equation}\label{30}
H (t) \simeq \frac{2}{3 t},\quad \rho(t) \simeq \frac{1}{6 \pi G t^{2}}.
\end{equation}
Choosing the constant $C \simeq 1 + \frac{\tau}{t_{0}} \ln 2$, for $t \gg t_{0} > 2 \tau$ we obtain the relations for the relativistic matter
\begin{equation}\label{31}
H (t) \simeq \frac{1}{2 t},\quad \rho(t) \simeq \frac{3}{32 \pi G t^{2}}.
\end{equation}
For times $t \ll t_{0}$ and $t_{0} > 2 \tau$, from Eq. (\ref{27}) it follows
\begin{equation}\label{32}
D (t) = 2 t + (C + 1) t_{0} - \tau \ln 2.
\end{equation}
Setting $C \simeq -1 + \frac{\tau}{t_{0}} \ln 2$, these expressions reduce to
\begin{equation}\label{33}
H (t) \simeq \frac{1}{t},\quad \rho(t) \simeq \frac{3}{8 \pi G t^{2}}.
\end{equation}
The equation for $\rho(t)$ has a form of Whitrow-Randall's relation \cite{Whi}.

\section{Discussion}
The equations (\ref{26}), (\ref{27}), and (\ref{29}) demonstrate how the Hubble expansion rate and the energy density
change with time from the inflationary phase of the universe's expansion, through the subsequent eras of an intensive energy transfer 
and pressure-free 
matter, to the beginning of the radiation domination. By choosing the constant of integration $C$, the solutions (\ref{26}) and (\ref{29})
are reduced to known `standard' expressions (\ref{30}), (\ref{31}), and (\ref{33}). An interesting feature of the solution (\ref{29}) 
is that at the point $t = 0$ it is finite,
\begin{equation}\label{34}
\rho(0) = \frac{3}{2 \pi G [(C + 1) t_{0} - \tau \ln 2]^{2}}.
\end{equation}
Thus, the two-component system does not have an initial cosmological singularity. 

The equations (\ref{26}), (\ref{29}), and (\ref{32}) can be 
continued into the region of extremely small values of time, $t \ll \frac{1}{2} |(C + 1) t_{0} - \tau \ln 2|$, 
where the Hubble expansion rate slightly changes with time, so that in the inflationary phase 
$H(t_{i}) \sim H(0) = \sqrt{\frac{8 \pi G}{3} \rho(0)}$ and the expansion of the universe
will be exponential in time, $R(t) \sim \exp \{H(t_{i}) t\}$.

The expression for the energy density $\rho (t_{i})$ in the inflationary phase can be reduced to the `standard' form. 
 Setting $G = M_{P}^{-2}$ \cite{Ly}, 
where $M_{P}$ is the Planck mass, $\tau \simeq M_{P}^{-1}$, and choosing $C = - 1$, from Eq. (\ref{34}) with a good accuracy we get
$H(t_{i}) \simeq \sqrt{\frac{8 \pi}{3}} M_{P}$, $\rho (t_{i}) \simeq M_{P}^{-4}$ (cf. \cite{Lin}).

In the two-component model (\ref{10}) with the equation of state (\ref{3}) with the parameter (\ref{9}), the evolution of the universe 
goes in one direction, from small times $t \ll t_{0}$ to large values $t \gg t_{0}$.

The following mechanical analogy allows to understand the reason of the origin of this `arrow of time'.
The function (\ref{9}) can be considered as the kink solution of the equation
\begin{equation}\label{35}
\frac{1}{2} \dot{w}^{2} + [- U(w)] = 0, \quad U = \frac{9}{2 \tau^{2}} \left(w^{2} - \frac{1}{9}\right)^{2},
\end{equation}
which describes the motion of the analogue particle with zero energy in the potential $[- U(w)]$ (cf., e.g., Ref. \cite{Raj}).
This potential has two maxima at the points $w = \pm \frac{1}{3}$ and a local minimum at $w = 0$. 
The analogue particle moves along the `trajectory' (\ref{9}) from the value $w = - \frac{1}{3}$ in the distant past ($t = - \infty$)
to the value $w = \frac{1}{3}$ reached at $t = \infty$. At the moment $t = t_{0}$, the particle passes through the minimum
of the potential at $w = 0$. Leaving the point $w = - \frac{1}{3}$, the analogue particle can only approach the point $w = \frac{1}{3}$
at $t \rightarrow \infty$, where its velocity and acceleration vanish. It cannot return back to $w = - \frac{1}{3}$. 

We note that Eq. (\ref{35}) has another solution in the form of the antikink which is equal to the function (\ref{9}) with an inverse sign.
This case corresponds to the model in which the relativistic matter at $t = - \infty$ transforms into the pressure-free matter and then
into a gas of low-velocity cosmic strings at $t = \infty$. It was studied in Ref. \cite{Kuz}, where it was shown that
the equation of state of matter can change with the expansion of the universe due to energy transfer between
the matter components (scalar fields) allowing to reproduce the evolution of matter in the universe with non-zero cosmological constant.
The description on equal footing of the universe over the total time interval from inflation through reheating to subsequent cooling
and transition to the pressure-free matter using the kink and antikink solutions of Eq. (\ref{35}) may indicate about their important role
in the representation of the evolution of matter.

\providecommand{\href}[2]{#2}\begingroup\raggedright\endgroup

\end{document}